\documentclass[useAMS, usenatbib]{mn2e}

\usepackage{xspace}
\usepackage{amsmath}
\usepackage{epsfig}
\usepackage{defs}

\def\ie{{\it i.e.}}

\def\ltsima{$\; \buildrel < \over \sim \;$}
\def\simlt{\lower.5ex\hbox{\ltsima}}
\def\gtsima{$\; \buildrel > \over \sim \;$}
\def\simgt{\lower.5ex\hbox{\gtsima}}

\def\hide#1{}

\title[Two Epochs of Globular Cluster Formation]{Two Epochs of Globular Cluster Formation from Deep Fields Luminosity Functions:
  Implications for Reionization and the Milky Way Satellites}

\author[H. Katz \& M. Ricotti]{Harley Katz$^{1}$\thanks{E-mail: hkatz@astro.umd.edu} and Massimo Ricotti$^{1}$\thanks{E-mail: ricotti@astro.umd.edu}\\
$^1$Department of Astronomy, University of Maryland, College Park, MD 20742, USA
} 

\begin{document}

\maketitle

\begin{abstract}
The ages of globular clusters in our own Milky Way are known with
precision of about $\pm 1$~Gyr, hence their formation history at
redshifts $z \simgt 3$ and their
role in hierarchical cosmology and the reionization of the intergalactic
medium remain relatively undetermined.  Here we analyze the effect of globular cluster formation on
the observed rest-frame UV luminosity functions (LFs) and UV continuum
slopes of high redshift galaxies in the Hubble Ultra Deep Fields. We
find that the majority of present day globular clusters have formed
during two distinct epochs: at redshifts $z \sim 2-3$ and at redshifts
$z\simgt 6$. The birth of proto-GC systems produce the steep, faint-end
slopes of the galaxy LFs and, because the brightness of proto-GCs fades $5$~Myrs
after their formation, their blue colors are in excellent agreement
with observations.

Our results suggest that: i) the bulk of the old globular cluster population
with estimated ages $\simgt 12$~Gyr (about $50\%$ of the total population),
formed in the relatively massive dwarf galaxies at redshifts $z\simgt
6$; ii) proto-GC formation was an important mode of star formation in
those dwarf galaxies, and likely dominated the reionization process.
Another consequence of this scenario is that some of the most massive
Milky Way satellites may be faint and yet undiscovered because tidal
stripping of a dominant GC population precedes significant stripping
of the dark matter halos of these satellites. This scenario may alleviate some remaining
tensions between CDM simulations and observations.
\end{abstract}

\begin{keywords}
(Galaxy:) globular clusters: general, galaxies: luminosity function, mass function, cosmology: theory
\end{keywords}

\section{Introduction}

Although globular clusters (GCs) are well studied in our local
galactic neighborhood, their role in hierarchical cosmology and their
formation history at redshifts $z>3$, when the universe was less than
$\sim 2$~Gyrs old, remains an unsolved problem.  Several groups have
attempted to model the formation of GCs using cosmological simulations
\citep[e.g.,][]{Griffenetal2010, Zonoozietal2011}, however,
insufficient spatial resolution and a poor understanding of their
initial mass function, as well as the mechanisms responsible for their
formation leave many unanswered questions about their cosmological origin.

Nearly all GCs are compact self-gravitating star clusters fairly
homogeneous in heavy elements \citep{Sneden2005}.  This suggests that
most of their stars formed in a nearly instantaneous burst with high
efficiency of gas-to-star conversion \citep{Jamesetal2004,
  Carrettaetal2009a}.  GCs are often classified into two categories:
metal-poor with [Fe/H]$\le -1.5$ and metal-rich with [Fe/H]$\ge
-1.5$. The origin of these two populations is unknown, however this
bimodal property has been observed in multiple galactic environments
\citep{ZepfAshman1993}.  The ages of many Milky Way GCs and a few
extra-galactic GCs have been estimated by analyzing the
color-magnitude diagrams of the individual GCs.  Vertical, $\Delta$V,
methods tend to have error bars of $\sim 1$~Gyr due to uncertainties
in defining the precise main sequence turnoff, while horizontal
methods, $\Delta$(B-V), have similar error bars as a result of
uncertainties in the theoretical model for the effective temperature
$T_{eff}$ of the stars \citep{Sarajedini2009}.  In addition, galactic
metal poor GCs are older than metal rich GCs and have an age spread of
approximately $1$~Gyr, compared to the $\sim 6$~Gyr spread for metal
rich GCs
\citep{Chaboyer1996,Rosenbergetal1999,SalarisWeiss2002,DeAngeli2005,
  Marinetal2009}. The age gap between metal-poor and metal-rich GCs is
greater than the range present within each individual population
\citep{Marinetal2009}, suggesting that there are two distinct epochs
of GC formation, adding to the complexity in modeling their formation
and evolution history.

\begin{figure*}
\centerline{\epsfig{figure=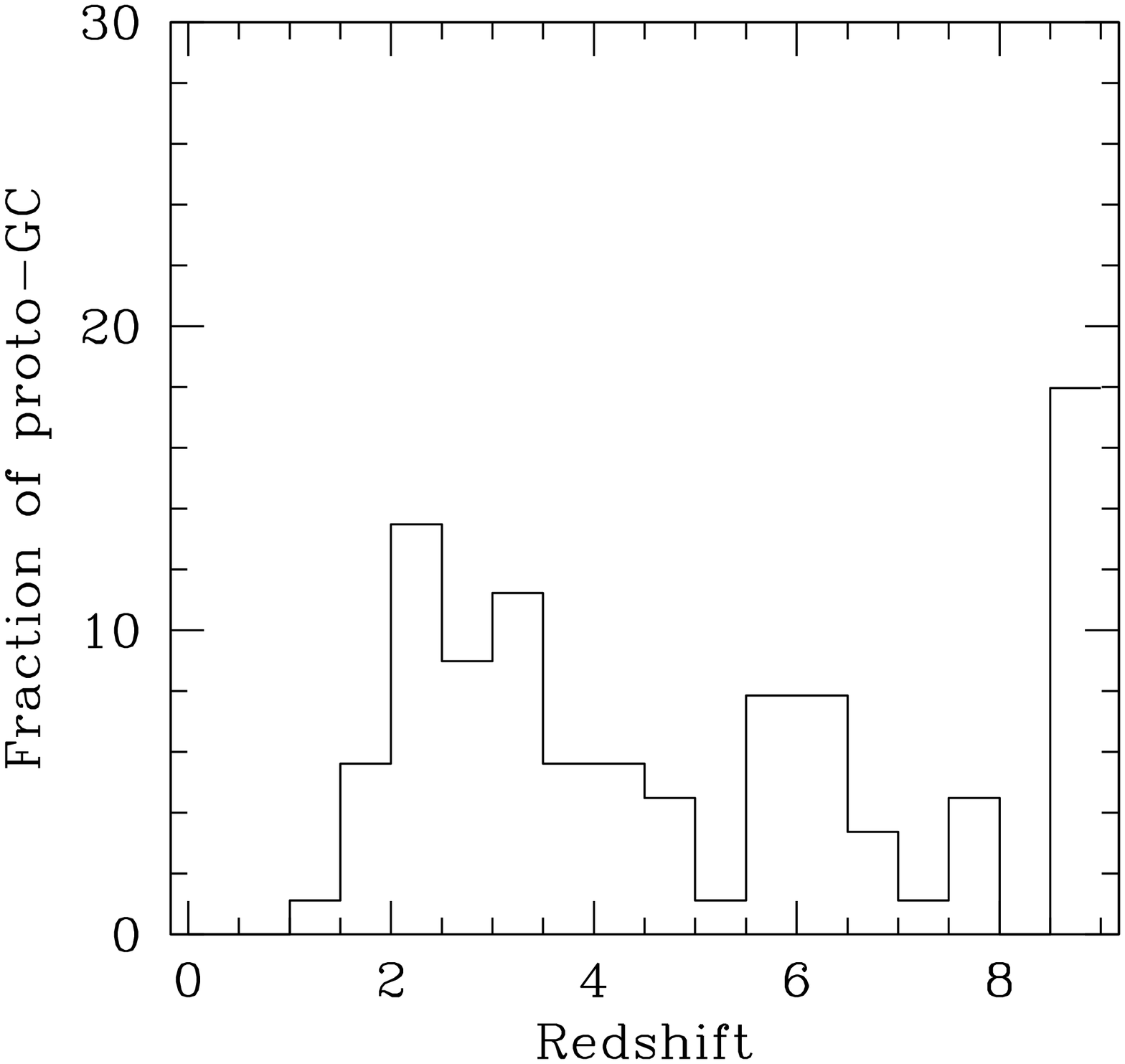, width=8.5cm}\epsfig{figure=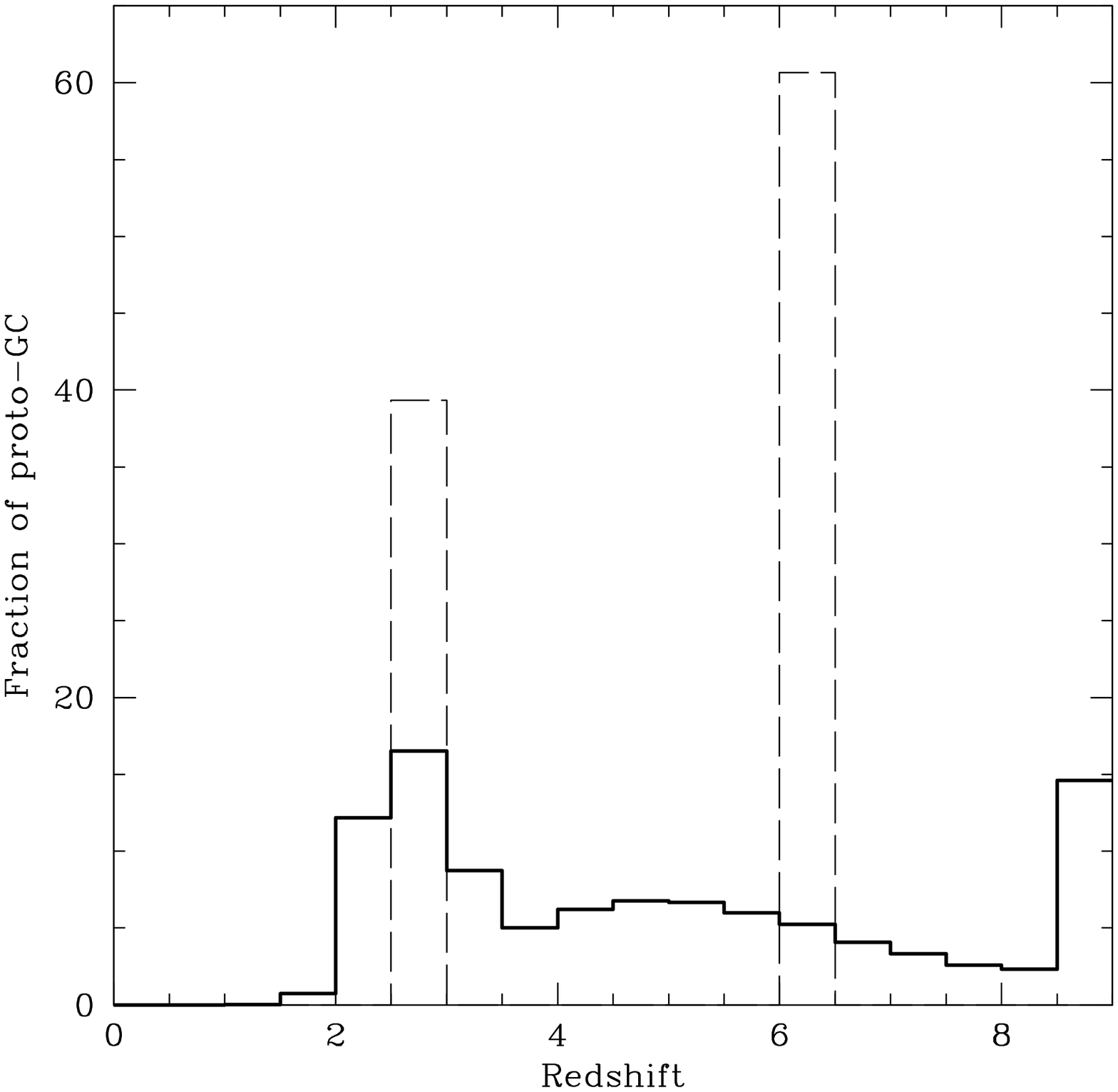, width=8.5cm}}
\caption{{\it Left.} Distribution of the redshift of formation of the
  Milky Way GCs based on age measurement from
  Forbes \& Bridges (2010). The distribution is poorly known at $z>3$
  because the typical error on the ages of GCs is $\pm
  0.5-1$~Gyr. {\it Right.} Monte Carlo realization of the redshift
  distribution (including errors on the ages) of a bimodal population
  of GCs that form 40\% of all clusters at $3<z<3.5$ and 60\% at
  $5<z<5.5$. Due to the large errors on the age measurements, it is
  impossible to discriminate between different choices for the
  underlying GC bimodal distribution. Most choices, as the one shown
  here, produce observed distributions broadly consistent with the
  Milky Way population of GCs.}
\label{fig:gc0}
\end{figure*}

Assuming that the bulk of the old GC population formed around the
epoch of reionization, \cite{Ricotti2002} has shown that they would
dominate the reionization process. Ricotti's argument is simple: from
the census of GCs in the present universe (estimated from the specific
frequency of GCs), he derived the fraction of baryons in GCs and the
number of ionizing photons per baryon per Hubble time they
emitted (assuming they formed over a 1~Gyr period at $z\sim 6$). This number is
likely to be a lower limit since most GCs have been destroyed or have
been stripped of stars due to dynamical processes and stellar
evolution \citep{OstrickerGnedin1997,FallZhang2001}. Today, the
fraction of stars in GCs today is negligible when compared to all of the
stars in bulges and disks. However, at redshift $z\simgt 6$, only a small
fraction of today's stars existed, while many of today's GCs, being
the oldest known stellar systems, may have been present. Ricotti
concluded that proto-GC formation may have been the dominant mode of
star formation in primordial dwarf galaxies. In addition, because
proto-GCs formed with a high star formation efficiency and are observed
in the outer parts of dark matter halos, the UV radiation they radiate
can more easily escape into the intergalactic medium (IGM)
\citep{RicottiS2000}, and contribute to the reionization
process. Indeed, assuming a Salpeter IMF and UV escape fraction
$f_{esc}\sim 1$, Ricotti showed that GC formation at $z \simgt 6$ can
easily produce enough UV photons to reionize the universe.

The main uncertainty in this model is the fraction of the observed
population of old GCs that formed at redshifts $z\simgt 6$.  The epoch
of reionization likely extended from $z \sim 14$ to $z \sim 7$
\citep{Fanetal2006}.  The uncertainty on the age determination of old
GCs prevents us from discriminating whether old GCs formed at redshift
$z \sim 3-4$, after the epoch or reionization, or at redshift $z \sim 7-14$, thus contributing to
reionization. In order to understand the role that GCs may have played
during the reionization epoch, the formation history of the old GC
population with estimated ages $\simgt 12 \pm 1$~Gyr (\ie, formation
redshift $z \simgt 4$) must be better constrained.  In addition,
constraining proto-GC formation at high redshift may be a key element
to understanding the origin of the classical and ultra-faint Local
Group dwarfs \citep{RicottiG2005, BovillR2011a, BovillR2011b, Boylan-Kolchin2011}.

Figure~\ref{fig:gc0} (left) shows the distribution of the redshift of
formation of the Milky Way GCs based on age measurement compiled by
\cite{ForbesBridges2010}. The distribution shows that about $55\%$ of
Milky Way GCs are older than 12~Gyrs, thus formed at redshift $z>4$.
However, the errors on these ages are on the order of $\pm 1$~Gyr.
Assuming the worst case scenario where all of the GCs are $1$~Gyr
younger than the data suggests, $20\%$ of Milky Way GCs would have
still formed after redshift $z>4$. The right panel of
Figure~\ref{fig:gc0} shows a Monte Carlo realization, consistent with
observations of the redshift distribution (including Gaussian errors
on the ages) of a bimodal population of GCs that form 20\% of all
clusters at $3<z<3.5$ and 80\% at $5<z<5.5$. Due to the large errors
on the age measurements, is impossible to discriminate between
different choices of the underlying GC bimodal distribution. For
instance, a bimodal distribution with 45\% of GC forming between
$2.5<z<3$ and 55\% at $6.5<z<7$, is statistically indistinguishable
from the one shown in the figure.

The old stellar populations and compact size of present day GCs adds
to the difficulty in detecting them beyond the Milky Way, even with
our most powerful space telescopes. \cite{Kaliraietal2008} has
identified bright unresolved
sources surrounding the parent galaxy at $z=0.0894$ in a Hubble space telescope (HST) image, which is one of
the furthest known systems of old GCs at a luminosity distance of
about $404$~Mpc.

For about $5-10$~Myrs after their formation, individual
proto-GCs are extremely bright and have blue broad-band colors. In
principle they are detectable with current instrumentation up to
redshift $z \sim 7$ (see Section~\ref{sec:idea}) and are certainly
within the reach of even deeper near-IR observations which will be
achieved with the James Webb space telescope (JWST)
\citep{SchaererCharbonnel2011}.  Moreover, GCs are typically observed
as systems of $3-10$ GCs in dwarf galaxies \citep{Puzia}, to hundreds
in more massive galaxies (the Milky Way has about $150$ GCs and
Andromeda about $500$).  With current space telescopes, proto-GC
systems forming in high redshift dwarf galaxies are poorly resolved
spatially. In Section~\ref{sec:images} we show that they appear as
blurred extended objects, brighter than an individual GC, since their
formation is likely nearly coeval given the typical dynamical time
scale of a few tens of Myrs in high-z dwarfs.  These young proto-GC
systems are therefore more easily detectable than sparsely distributed
individual GCs in larger galaxies.

In this paper, we take an indirect approach to identifying the
redshift of formation of GCs. We seek to constrain the largest number
of proto-GCs that are allowed to form within any given redshift range,
while remaining consistent with the observed UV luminosity functions
(LFs) and colors of galaxies in the Hubble Ultra Deep Fields. We also
produce simulated images of young proto-GC systems to lay the
foundation for their direct identification with Wide Field Camera 3
(WFC3) on the HST and the Near Infrared Camera on the JWST.

This paper is organized as follows.  In Section~\ref{sec:model} we
present basic calculations of the luminosity and mass budget of
proto-GCs. In Section~\ref{sec:idea}, we discuss the capabilities of
the HST and the JWST in detecting and resolving young proto-GCs and
present synthetic images of proto-GC systems based on observations of
old GC systems in nearby dwarf galaxies by \cite{Puzia}.  In
Section~\ref{sec:LF}, we constrain the proto-GC formation rate at
high-z by comparing observed LFs to simulated LFs of forming
proto-GCs, and calculating the effect of proto-GCs formation on the UV
continuum slope at $1500$~\AA.  We present our
results and discussion in Section~\ref{sec:res} and the summary and
conclusions in Section~\ref{sec:conc} .

\section{Basic Calculations and Assumptions}\label{sec:model}

The mass distribution of present day GCs in the Milky Way is well
described by a lognormal distribution with typical mass $M_{\rm obs} \sim
1.4 \times 10^5$~M$_\odot$ \citep{Shuetal2010}. However, the initial
masses of young proto-GCs, $M_{\rm ini}$, were significantly larger than
today's observed masses, $M_{\rm obs}$.  Stellar evolution and dynamical
effects, including galactic tides, dynamical friction and two-body
relaxation \citep{OstrickerGnedin1997, FallZhang2001}, significantly
reduce the number and mean mass of young GCs from the epoch of
formation to the present.  Assuming a Salpeter initial mass function
(IMF), \cite{PrietoGnedin2008} found $f_{\rm di} \equiv M_{\rm ini}/M_{\rm obs}
\sim 9.1$. In their simulations they include the effects of stellar
evolution, two-body relaxation, dynamical friction and tidal shocks as
described by \cite{FallZhang2001}. Simulations also show that the
current mass distribution of GCs can be reproduced if the GC initial
mass function (GCIMF) is either a power-law $dN/dlogM \propto M_{\rm ini}^{-0.47}$
with $10^4<M_{\rm ini}<10^7$~M$_\odot$ or a lognormal distribution with
characteristic mass $M_{\rm ini} = 1.5 \times 10^6$~M$_\odot$ \citep{Shuetal2010}.

Although the destruction rate and mass loss of GCs remains somewhat
uncertain because these quantities are based on theoretical models
with poorly known initial conditions (such as the IMF of stars in GCs,
the GCIMF, and the formation sites of proto-GCs), recent observational
developments on the stellar populations in GCs provide independent
evidence for values of $f_{\rm di}\sim 10-25$
\citep{SchaererCharbonnel2011}.  It is now clear that nearly all GCs
exhibit large star-to-star variations of abundances in light elements
(from C to Al), interpreted as a signature of some degree of
self-enrichment by gas pre-processed in short-lived, high mass stars
\citep[for details see,][and references therein]{Grattonetal2001,
  Grattonetal2004, Prantzosetal2007, Carrettaetal2009b,
  Charbonnel2010}. Self-enrichment models of GCs assume at least two
populations of stars, with the first generation of massive stars
polluting the second generation. The polluters are thought to be
either massive AGB stars or fast rotating massive stars
\citep{PrantzosCharbonnel2006}.  These models, after assuming an IMF
for the first generation stars, are constrained to reproduce the
present proportion of first to second generation stars and their
observed abundance patters. Generally, all the models require the
initial stellar masses of GCs to be considerably larger than their
present day values \citep{PrantzosCharbonnel2006, Decressinetal2007,
  DErcoleetal2008, Decressinetal2010, Carrettaetal2010}.
\cite{SchaererCharbonnel2011}, using the model of \cite{Decressinetal2007}, have determined $f_{\rm di}$ as a function of the IMF of the
first generation of stars. They found $f_{\rm di}\sim 8-10$ in models in
which the second generation stars are all retained by the GCs, and
$f_{\rm di} \sim 15-25$ in models in which second generation stars found
in the Milky Way halo originate from the present-day population of
GCs. This result is independent of, but in good agreement with the
expectation from dynamical models \citep{PrietoGnedin2008}. Thus, for
the rest of this paper we will adopt a conservative fiducial value
$f_{\rm di} = 9.1$, at the lower end of the estimate by
\cite{SchaererCharbonnel2011}. Note however that our results on
constraining the formation rate of proto-GC systems do not depend on
the assumed fiducial value for $f_{\rm di}$.

\subsection{Luminosity Function of proto-GCs}
\begin{figure*}
\centerline{\epsfig{figure=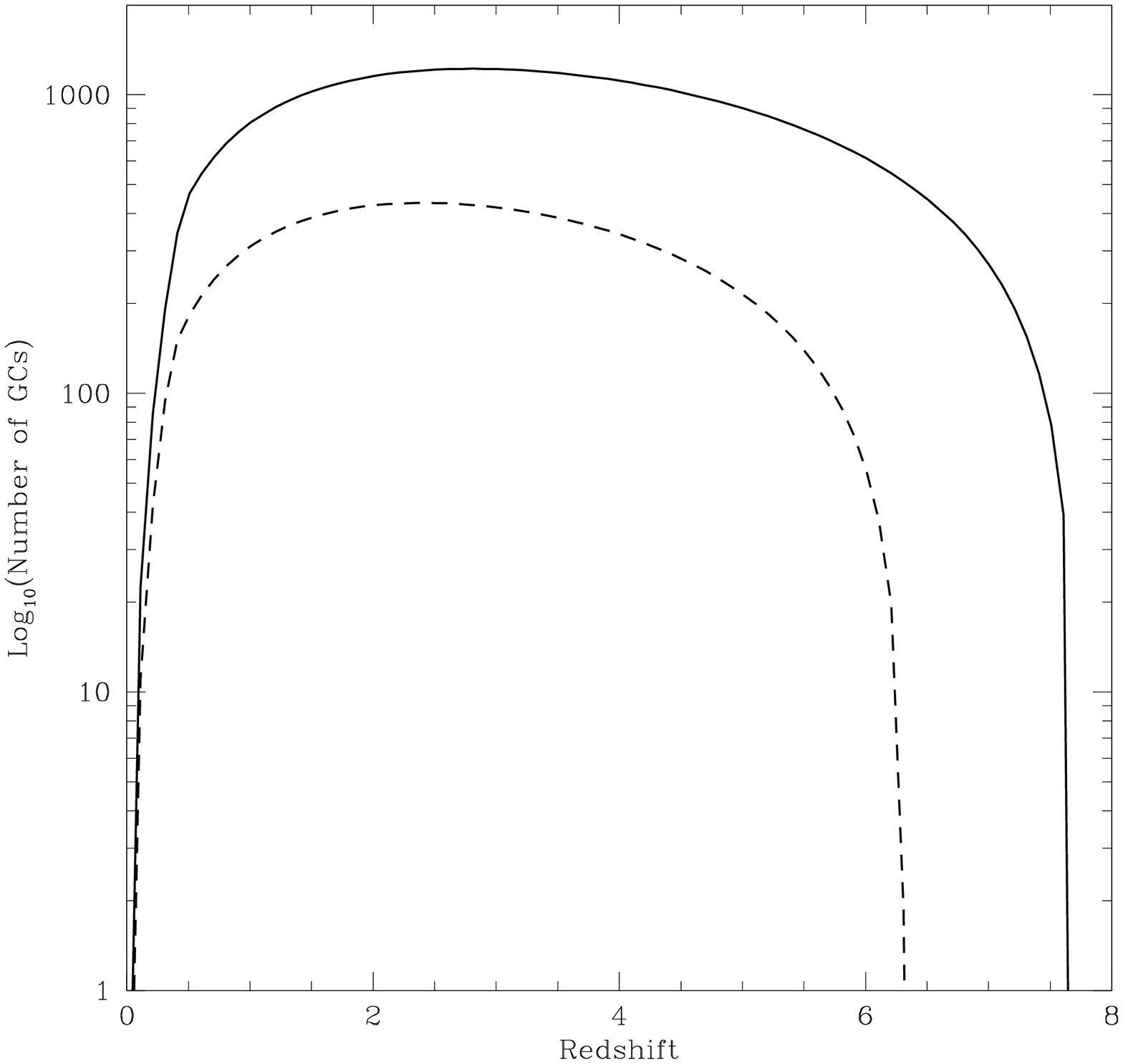, width=8.5cm}\epsfig{figure=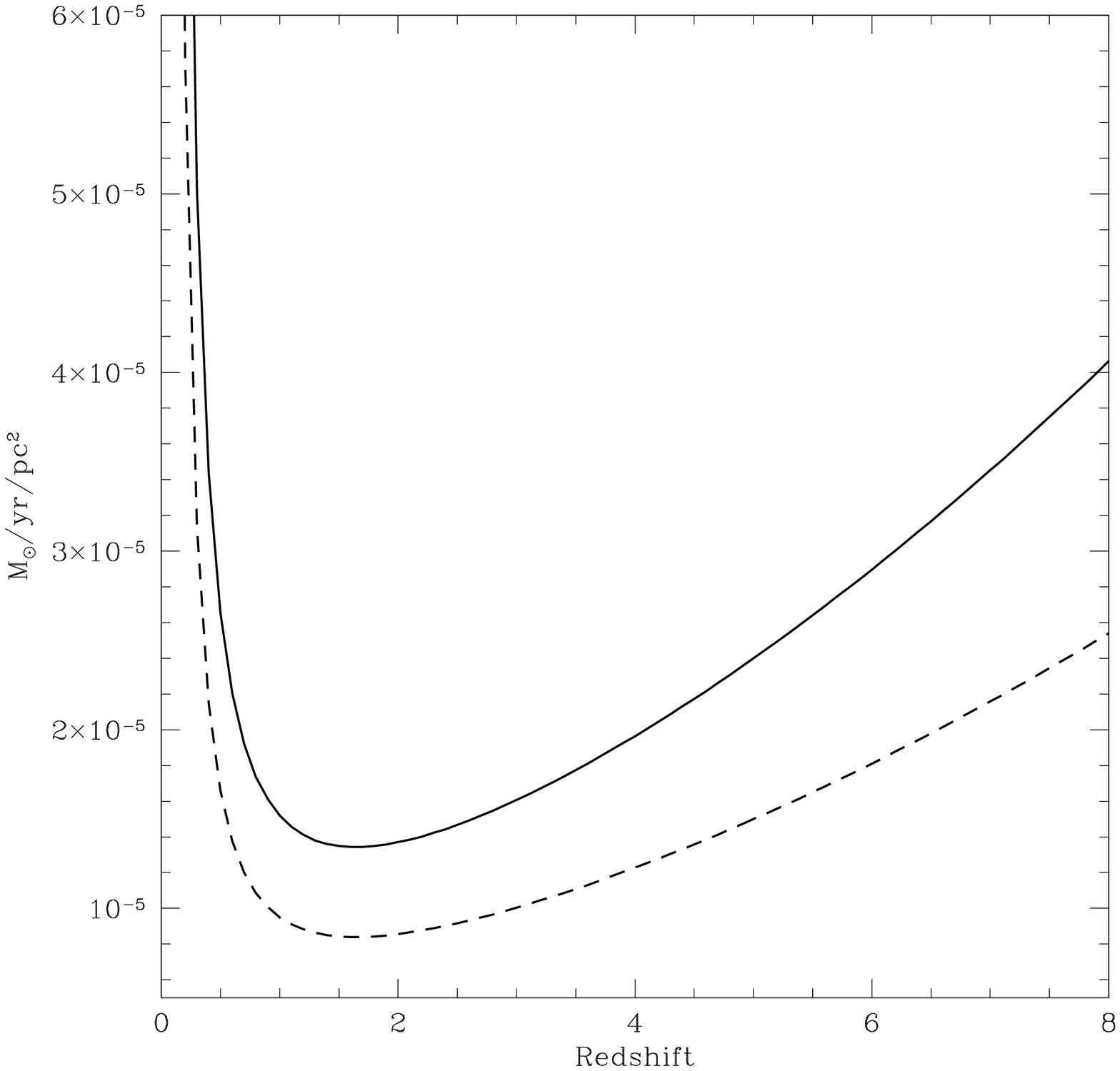, width=8.5cm}}
\caption{{\it Left}. The solid black line represents the average
  amount of GCs that the NIR camera on JWST would see as a function of
  redshift.  The dashed line represents the average amount of GCs that
  the WFC on HST would see as a function of redshift.  This assumes a
  limiting magnitude of 31 for JWST and 30.5 for HST {\it Right}. The
  star formation rate per square parsec as observed with JWST is shown
  as the solid black line and the dashed line represents the same quantity
  as observed with HST.}
\label{fig:gc1}
\end{figure*}

For a lognormal GCIMF, with a typical mass of $M_{ini} =1.4 \times
10^6$~M$_\odot$, a young proto-GC is expected to have a peak UV
magnitude at $1500$~\AA\ of $M_{AB}\approx -16.7$ at an age
$<5-10$~Myr. These GCs then fade by more than 4 mags due to stellar
evolution (see e.g. models of \cite{Leitherer1999}). For a power law GCIMF, the most massive GCs have $M_{ini}^{max}
\sim 10^7$~M$_\odot$ \citep{Shuetal2010}, corresponding to $M_{AB}
\approx -19.3$.  At redshifts of $z \sim 7-10$ this would correspond
to a typical UV rest frame magnitude of $m_{AB} \sim 30.3$ at the peak
of brightness, slightly fainter than the current detection limits of
the deepest near-IR images taken with the WFC3 camera onboard of the
HST \citep{Bouwensetal2010}.

\subsection{Globular Cluster Formation Rate and Mass Budget}

Using published data on the specific frequency of GCs in spiral,
elliptical and dwarf galaxies, \cite{Ricotti2002} estimated that about
$\omega_{gc} \sim f_{di}(2.7 \times 10^{-4}) = 0.25\%$ of cosmic
baryons have been converted into GC stars. Although this fraction is
about 40 times smaller than today's baryon fraction in stars (about
$10\%$), it is likely a significant fraction of the collapsed baryons
at redshifts $z \simgt 6$.  Based on a compilation of observations of
the present number of GCs in galaxies, \cite{Puzia} estimate an
efficiency of GC formation in dark matter halos, $\eta_{dm} \sim 5.5
\times 10^{-5}$, defined as $M_{gc}^{\rm tot} \equiv \eta_{dm} M_{dm}$
(assuming standard cosmological parameters $\Omega_b=0.04$ and
$\Omega_{dm}=0.21$). Since at $z=0$ most dark matter is in collapsed
halos, the results of \cite{Puzia} imply $\omega_{gc} \equiv \eta_{dm}
\Omega_b/\Omega_{dm} \sim 2.9 \times 10^{-4}f_{di}$, which is very good agreement with the value estimated by \cite{Ricotti2002}.  Hereafter we will
adopt a fiducial value for the mass density of proto-GCs at formation:
\begin{equation}
\rho_{gc} \equiv \omega_{gc} \overline{\rho_b} \sim 1.35 \times
10^7~{\rm M}_\odot~{\rm Mpc}^{-3},
\end{equation}
where we have assumed $f_{di}=9.1$ and the mean cosmic baryon density
$\overline{\rho_b}=5.51 \times 10^9$~M$_{\odot}$Mpc$^{-3}$
\citep{Komatsuetal2011}.

It is easy to show that if even a small fraction $f_{gc} \sim 20\%$ of
today's GCs formed at redshift $z>7$ (over a time interval $\Delta t
=0.5$~Gyr), the GC formation rate per comoving volume would be
\begin{equation}
\dot \rho_{gc} \approx f_{gc} \frac{\rho_{gc}}{\Delta t}\approx 
0.54 \times 10^{-2} \left(\frac{f_{gc}}{20\%}\right)~{\rm M}_\odot yr^{-1}~{\rm Mpc}^{-3}, 
\end{equation}
comparable to the star formation rate density observed at redshift $z
\sim 8$ \citep{Bouwensetal2011}. These simple estimates suggest that
GC formation can be an important mode of star formation at high-z, and
perhaps dominate the reionization process.

\section{Detecting Proto-globular Cluster Systems in Hubble Deep Fields}\label{sec:idea}

The planned launch of the JWST provides exciting prospects for our
understanding of the high redshift universe.  Here, assuming
negligible dust extinction, we address the capabilities of both the
HST and the JWST to detect young proto-GCs at high redshift. First, we
estimate the absolute luminosity, number and morphology of young
proto-GC systems in HST deep fields and make predictions for JWST deep
fields.

The left panel in Figure~\ref{fig:gc1} shows the number of
GCs, $N_{gc}$, in the field of view of the Wide Field Camera
(WFC) on the HST (dashed line) and the Near Infrared Camera (NIR) on
the JWST (solid line). For illustration purposes, we have assumed that
half of all proto-GCs (after correcting for destruction), $f_{gc}=50\%$,
have formed in a single episode at redshift $z$. The value of $N_{gc}$ is estimated as
\begin{equation}
N_{gc}(z)=3\theta_An_{gc}f_{lim}(z)\frac{D_{L}(z)}{1+z},
\end{equation}
where $\theta_A$ is the field of view of the camera, $D_L(z)$ is the
luminosity distance at the redshift of interest, $f_{lim}(z)$ is the
fraction of proto-GCs brighter than the limiting magnitude of the
telescope at that redshift, and $n_{gc}$ is the number density of GCs
($n_{gc}\approx 39.1$~Mpc$^{-3}$ for a power law IMF and
$n_{gc}\approx 33.8$~ Mpc$^{-3}$ for a lognormal IMF).  Values of
$N_{gc}$ for different $f_{gc}$ can be easily inferred by scaling the
plot in Figure~\ref{fig:gc1} (left).  In principle, the WFC3 on the
HST has the capabilities to detect proto-GCs with $M_{ini} =
10^7~M_\odot$ at redshifts up to $z=7$ (see Section~\ref{sec:images}).
However, such high mass clusters are only $f_{mag} \sim 0.8\%$
of the total \citep{Shuetal2010}.  $N_{gc}$ peaks at $z \sim 7-8$
because although the number of proto-GCs increases with redshift due
to the larger volume in the field of view, the limiting magnitude of
the instrument reduces the fraction of GCs that are detectable to the
few most massive ones. In conclusion, single proto-GCs during their
youth are detectable with current instrumentation up to $z \sim 6$, and
are certainly well within the reach of deeper near-IR observations
which will be achieved with the JWST. Although we cannot spatially
resolve these sources at high redshifts, there presence in WFC and NIR
images may be significant when compared to the number of high-redshift
galaxy candidates.

\subsection{GCs Versus Host Galaxy Surface Brightness}
\begin{figure*}
\centerline{\epsfig{file=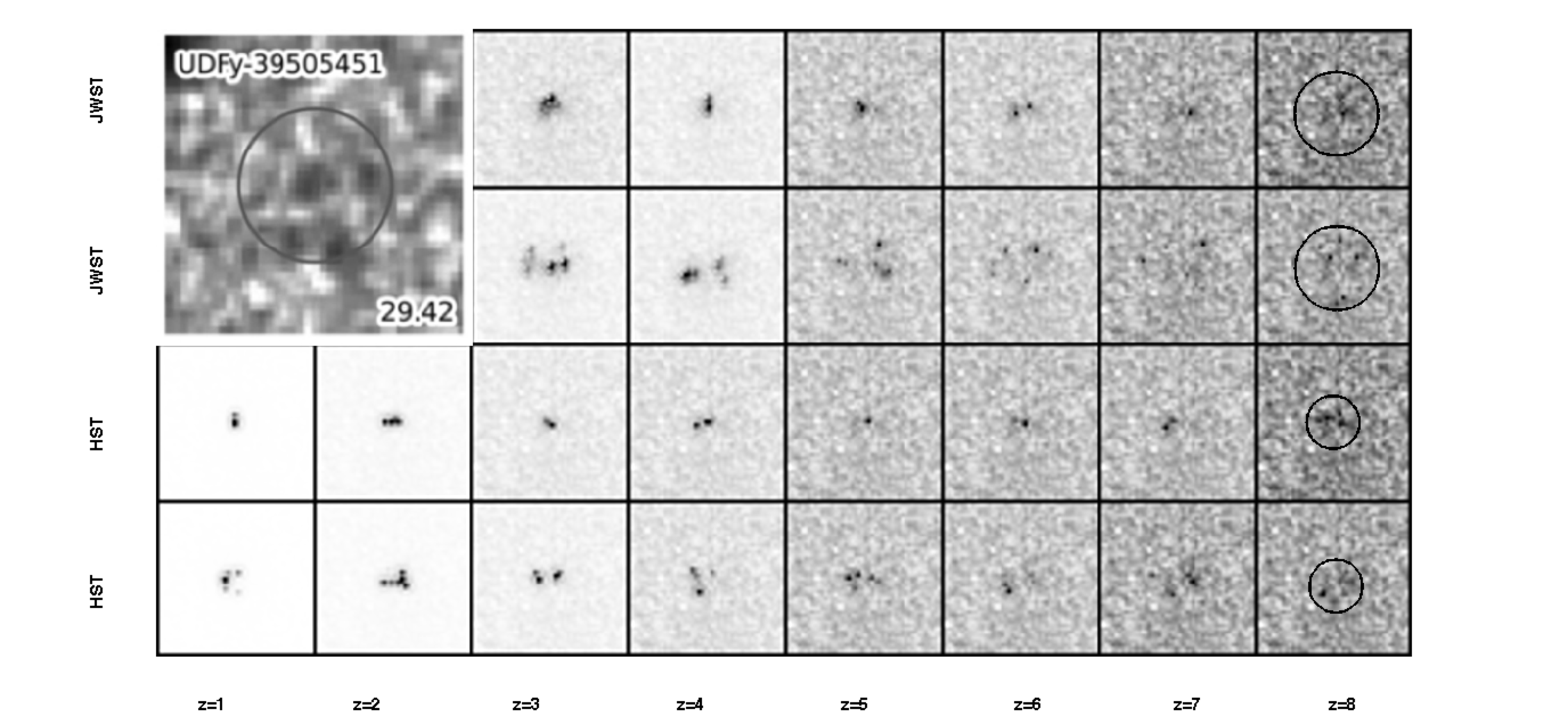, width=3.5cm}}
\caption{The top rows are the theoretical images from the WFC on HST
  and the bottom rows are the theoretical images from the NIRcam on
  JWST.  These were computed to model a total exposure time of
  1,000,000 seconds to be consistent with the HUDF.  HST sensitivity
  extends from $z=0-8$ from the UV and IR imaging equipment, while JWST
  only has sensitivity higher than $z=2.33$ for $1500$~\AA.  At the
  higher redshift we can see that the individual GCs are more
  identifiable.  PSFs for JWST were found in the simulated PSF library
  provided by STScI
  while the HST PSFs were generated with the Tiny Tim
  software.
  Shown at the top left is an actual image from
  Bouwens et al. (2011).  We use this to show a comparison of an
  actual $z=8$ candidate that appears to be multiple sources with the
  multiple source images that we have generated.}
\label{fig:gc2}
\end{figure*}

Using the average half light radius calculated from the Fornax Cluster
Survey VII \citep{Mastersetal2010}, of $2.8 \pm 0.3$ pc for red GCs
and $3.4\pm0.3$ pc for blue GCs, the maximum redshifts that WFC and
NIRcam can resolve these systems are $z=0.0028$ ($11.9$~Mpc) and
$z=0.0072$ ($30.7$~Mpc), respectively.  Without adaptive optics from
the ground it is impossible to resolve or take a spectrum of
GCs at intermediate and high redshifts. Thus, because of their high
star formation rate at formation, GCs will appear as especially bright
point sources surrounding their host galaxies and for about 5-10 Myrs
from formation they may outshine their host galaxies.  A GC of
$10^7~M_{\odot}$, with a half light radius of $3.4$~pc that forms all of
its stars in $\sim 10$~Myr, will have a SFR per unit area of ${\dot
  \Sigma_{gc}} \approx 2.75 \times
10^{-2}~M_{\odot}$yr$^{-1}$pc$^{-2}$.  However, since the proto-GC is
unresolved the effective SFR per unit area depends on the pixel size of
the detector, and at redshift $z=6$ is $\dot \Sigma_{gc}^{eff} \approx
1.81 \times 10^{-5}~M_{\odot}$yr$^{-1}$pc$^{-2}$ for the WFC on the
HST. As a comparison, the bulge of the Milky Way, which has a mass of
approximately $2 \times 10^{10}~M_{\odot}$ within a radius $\sim
3$~kpc, assuming its stars forms over $1$~Gyr, would have a SFR per
unit area of $\dot \Sigma_* \approx 2.83 \times
10^{-6}$~M$_{\odot}$yr$^{-1}$pc$^{-2}$. The Large Magellanic Cloud
(which we are using to compare GCs to dwarf galaxies) has a mass of $3
\times 10^{10}$~M$_{\odot}$ with a effective radius of $5$~kpc.
Assuming a continuous SFR over $1$~Gyr, we infer a SFR per unit area of
$\dot \Sigma_* \approx 3.82 \times
10^{-7}$~M$_{\odot}$yr$^{-1}$pc$^{-2}$.  Figure~\ref{fig:gc1} (right)
shows the effective SFR per unit area of forming proto-GCs as a
function of redshift and for two different proto-GC masses.  Because
proto-GCs are unresolved, with increasing redshift their effective
surface brightness increases due to the increase of the angular
resolution. Assuming that proto-GCs systems are poorly spatially
resolved and that their formation is synchronized to within 10~Myr as
described later, we expect that the light from a high-z dwarf galaxy
will be dominated by the proto-GC system for about 10~Myr after their
formation.

\subsection{Simulated Images}\label{sec:images}

The best constraint on proto-GC formation at high redshift is their
direct detection. As discussed in Section~\ref{sec:idea}, the large
star formation rates and compact sizes of proto-GCs allow them to
outshine their host galaxies for a brief time.  Figure~\ref{fig:gc2}
shows simulated images of proto-GC systems for 1~Msec observations
with HST and JWST at different redshifts, and for two different
choices of the host galaxy mass.  To generate the images, we have used
data on old GC systems observed in a survey of nearby spheroidal and
irregular dwarf galaxies \citep{Puzia}. We randomly generate systems
of GCs around these types of dwarf galaxies, modeled after the number
of GCs, and the magnitude and structural data of the host galaxy.  We
randomly select a radius, determined by averaging the volumes of shells
in which GCs orbit around eight different dwarf irregular galaxies. We
then project on the sky the GC position.  The magnitude of each GC is
also randomly assigned in a range of three magnitude bins from
$-18<M_{AB}<-15$.  We use the theoretical point spread function to
convolve the point source over many pixels.  Next, the
three-dimensional positions of these systems were projected onto a two
dimensional image with an arbitrary viewpoint and applied a normalized
gray scale that attributes relative brightness to the proto-GCs.  We
include the background noise with signal to noise ratios of the 1~Msec
exposure on each camera (consistent with that of the HUDF).

The brightness of the most massive proto-GCs is at the detection limit
for the two space telescopes.  Because of this, the signal to noise
ratio of these systems is low and it might often be impossible to
extract these systems over the background noise.  Additionally, the
theoretical images demonstrate that at around $z=1$, the systems of
GCs tend to produce an extended source rather than be resolved as
distinct objects.  This may effect the observers ability to
differentiate between GC systems and the host galaxy all the way up to
$z=4$.  Furthermore, if the systems form in groups rather than
individually, this effect may be amplified and further limit the
ability of an observer to identify GC systems.  However, at $z>4$,
this effect will be significantly reduced and the probability of
observing individual GCs or GC's in groups is far greater.

\section{Constraining the proto-GC formation rate at high-z with UV Luminosity Functions and Continuum Slopes}\label{sec:LF}

In this section we calculate the effect of the formation of proto-GCs
on the observed high-z UV luminosity functions (LF). The basic idea is
to build synthetic luminosity functions, assuming that the GC system
outshines the other stellar populations in the galaxy during the first
5-10~Myr after its formation. We use data from
\cite{Bouwensetal2011} for LFs at redshifts $z>7$,
\cite{Bouwensetal2007} for the B, V, and I dropout LFs, and
\cite{Oeschetal2010} for the $z= 1.5, 1.9$, and $2.5$ LFs.

\subsection{Unresolved proto-GCs}\label{sec:single}

Our first approach is to consider the formation of individual
proto-GCs. Although not the most realistic, this case is the simplest
and nearly model independent. Physically, this approach corresponds to
a scenario in which the formation of the proto-GC system in a dark
matter halo is not synchronized to within a few tens of Myrs, or the
most luminous proto-GC outshines all the others. This is the most
conservative assumption but is less appropriate at redshifts $z\simgt
6$ when the typical dynamical times in virialized halos are short:
\begin{equation}
t_{dyn} \equiv \frac{R_{s}}{v_{vir}}\approx 11~Myr \left(\frac{1+z_{vir}}{10}\right)^{-1.5},
\end{equation}
where $R_{s} \simeq R_{vir}/5$ and $v_{vir}$ are the scale radius and
the virial velocity of the halo, respectively.  We have adopted two
choices of the initial mass functions of proto-GCs (GCIMF): a
lognormal and power-law (see \S~\ref{sec:model}), both reproducing the present
day GC mass function in the Milky Way \citep{Shuetal2010}. Both
choices produce nearly identical constraints.

In order to calculate the proto-GCs LFs, as a fist order approximation
we assume that GCs form in a short burst with $\Delta t_{form}\simlt
10$~Myr (instantaneous mode of star formation).  Using Starburst-99
\citep{Leitherer1999} we derive the evolution of the absolute
magnitude of proto-GCs and their color (parameterized by $\beta$, the
log-slope of the spectrum at $1500$~\AA) as a function of time from
the burst. Proto-GCs produce the majority of their radiation in the UV
band in the first $5-10$~Myr of their life, before rapidly evolving to
higher magnitudes.  We compute the fraction $p_{gc}(M_{gc})$ of GCs
that form in each log mass bin, using constant log bin spacing of
$0.2$~M$_{\odot}$, spanning from $10^4$~M$_{\odot}$ to
$10^{7.2}$~M$_{\odot}$ for the power-law GCIMF and from
$10^{3.2}$~M$_{\odot}$ to $10^{7.4}$~M$_{\odot}$ for the lognormal
GCIMF.
The number of GCs at peak luminosity per unit comoving volume is
$n_{gc}(M_{gc})=f_{gc}f_{on}\rho_{gc}p_{gc}/M_{gc}$, where
$f_{gc}(z)$, the fraction of proto-GCs forming in each redshift bin,
is the free parameter we wish to constrain. Here, $f_{on}$ is the
fraction of time the proto-GC is at peak luminosity which is about
$f_{\rm on} \equiv 10~Myr/\delta T_{z}$, where $\delta T_{z}$ is the
time interval corresponding to the characteristic redshift depth of the
observed luminosity function. Finally, we compute the luminosity
function $\phi_{gc}$ from the proto-GC mass function converting from
stellar masses to magnitudes (using Starburst-99 for an instantaneous
burst).

In order to account for the evolution of GCs to higher magnitudes as
they age, we apply a correction to the calculation of the
LFs. Proto-GCs after the first 5-10~Myrs become fainter and redder,
but their number in each redshift bin is larger because they remain at
fainter magnitudes a longer time. We use Starburst-99 to determine the
time $\Delta t$ a proto-GCs of constant mass has a given
magnitude. The number of GCs at fainter magnitudes from the peak is
obtained analogously to the calculation above but with $f_{on}=\Delta
t/ \Delta T_{z}$. Thus, after converting proto-GCs masses to
magnitudes, we integrate over a time interval $\Delta T_{z}$.

\begin{figure*}
\centerline{\epsfig{figure=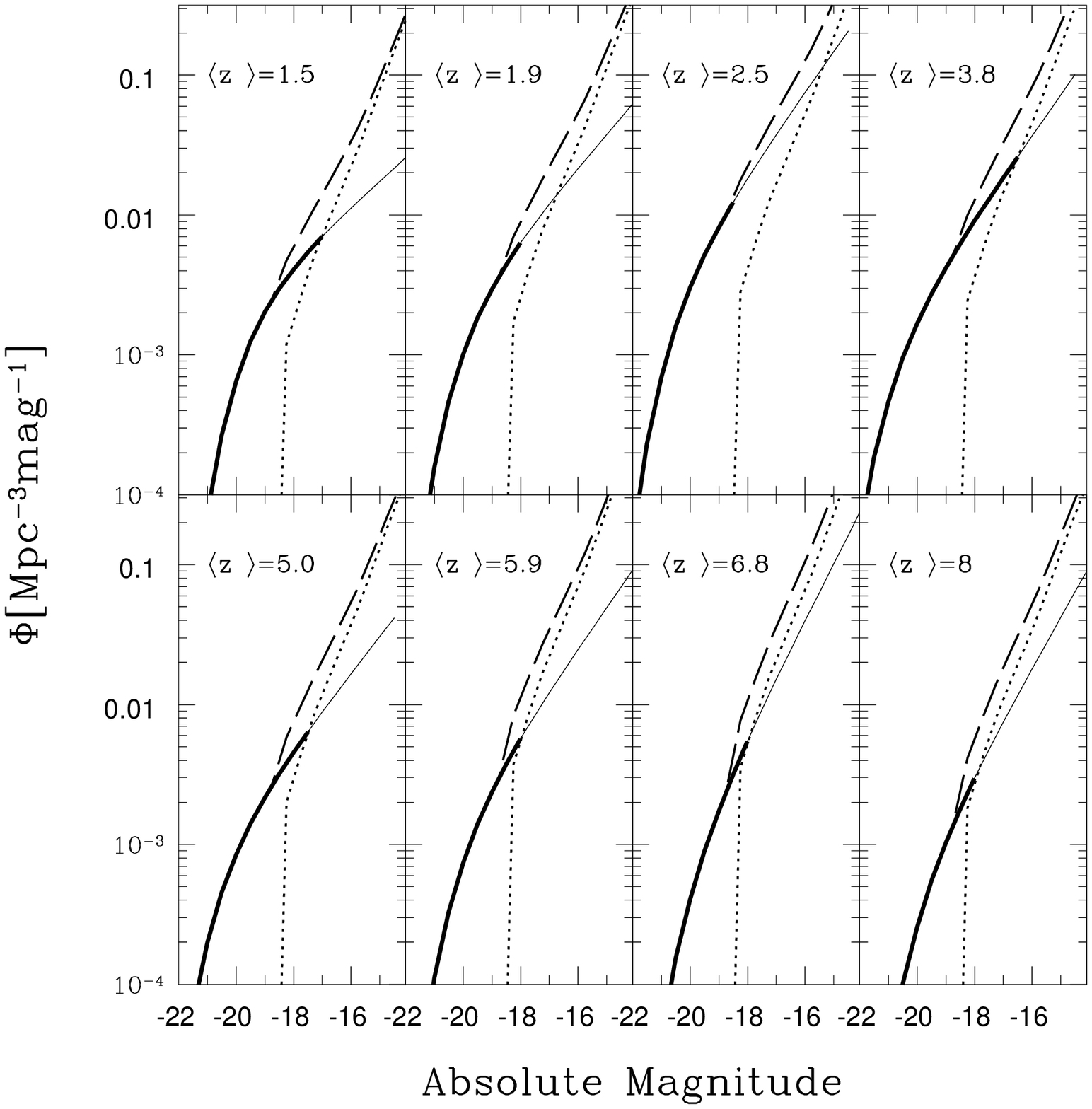, width=8.5cm}\epsfig{figure=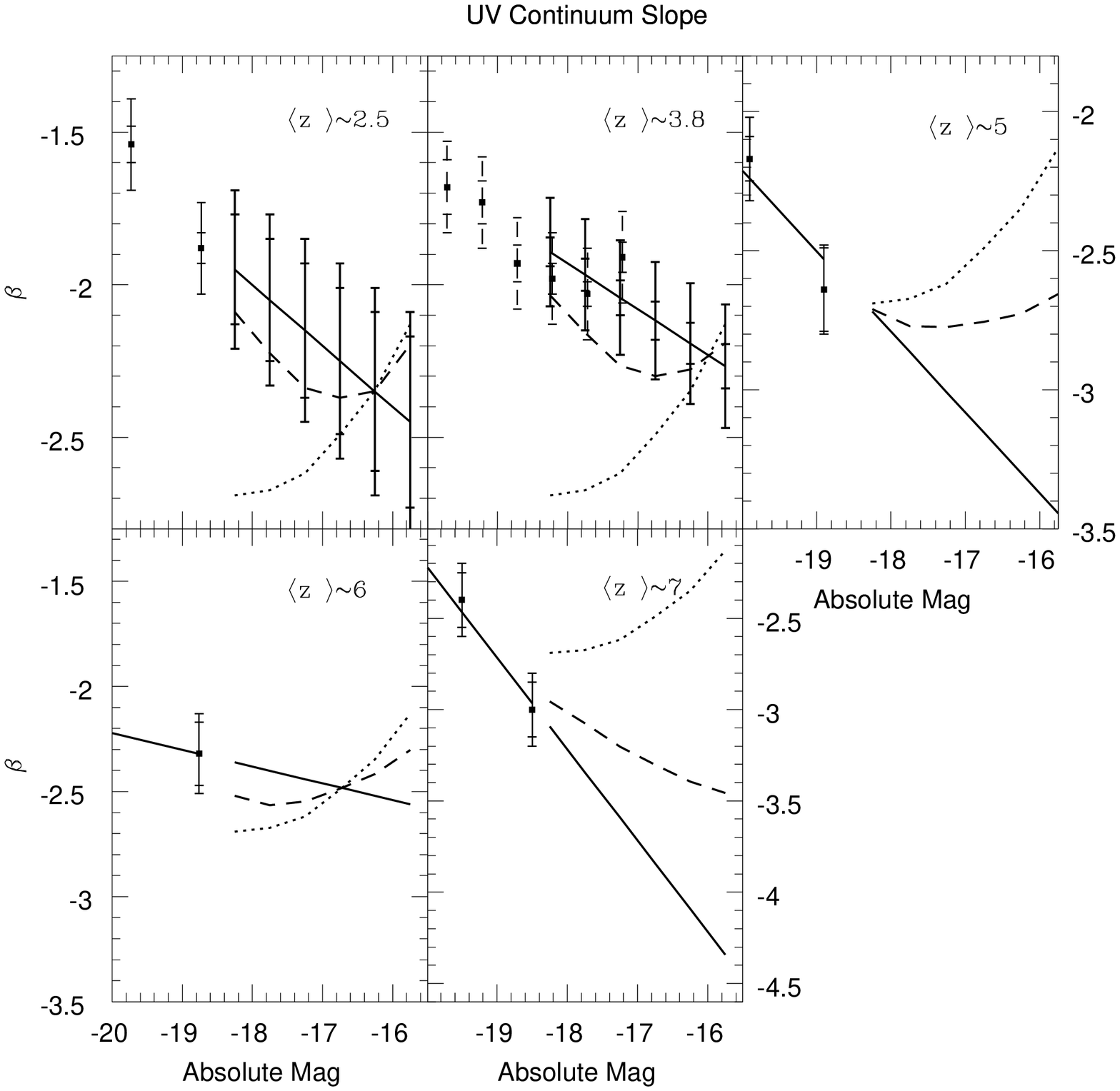, width=8.5cm}}
\caption{{\it Left.} The UV Luminosity Functions from
  Oesch et al. (2010) ($z=1.5, 1.9, 2.5$), Bouwens et al. (2007)
  ($z=3.8, 5.0, 5.9$), and Bouwens et al. (2011) ($z=6.8, 8$) are
  shown as the thick black line.  The observed luminosity function is
  only shown to the limiting magnitude of the observational data.  The
  dotted lines show the LF of a fraction $f_{gc}$ of proto-GCs forming
  in the same redshift bin as the observations, assuming a power law
  GCIMF.  The long-dashed line represents the sum of the proto-GCs
  LFs and observed LFs. {\it Right.} The UV continuum slopes for the
  individual GC formation model.  The observational UV continuum
  slopes are shown with the solid lines as the least squares slopes.
  The least squares fits are only shown with error bars if they were
  provided by the literature and extrapolations are shown with a gap.
  Raw data points and accompanying error bars are also shown from
  Bouwens et al. (2009) and Bouwens et al. (2011).
  The dotted lines are the
  theoretical UV continuum slopes for forming GCs and the dashed lines
  are the $\beta_{eff}$ slopes.}
\label{fig:gc3}
\end{figure*}

Figure~\ref{fig:gc3} (left) shows the UV LFs of (individual) proto-GCs
(dotted lines) compared to the observed LFs in in the HUDF at
different redshift intervals. The normalization of the proto-GCs LF in
each redshift bin is proportional to the parameter $f_{gc}$, that we
want to constrain. The proto-GCs LFs have steeper faint-end slopes
than the observed LFs, hence we are able to set upper limits on
$f_{gc}$. The values of $f_{gc}$ for each redshift bin are reported in
Table~\ref{tab:one}.

\subsubsection{Effective UV Continuum Slopes}\label{sec:slopes}

Once we have constrained $f_{gc}$ using the observed LFs, we analyze the effect of GC formation on the UV continuum slope at
$1500$~\AA.  Proto-GCs with ages $<5-10$~Myr have UV continuum slope
$\beta=-2.69$ \citep{Leitherer1999}.  We evolve the higher mass GCs into the lower
luminosity bins, as was described when comparing the luminosity
functions, and compute the weighted average of the $\beta$ at half
magnitude steps with respect to an arbitrary initial absolute
magnitude.  To determine the effective color, we set
$\beta_{eff}=f(\beta_{gc}-\beta)+\beta$, $\beta_{gc}$ is the color
parameter from the GCs and $\beta$ are the observed mean values of
deep field galaxies described by \cite{Bouwensetal2009}, and $f$ is
the fractional influence of the GCs on the UV continuum slope.  In
order to constrain the maximum percentage of GC formation, we first
measure $f$, the ratio of the proto-GC LF $\phi_{gc}$ to the observed
LF $\phi_{gal}$.  We then apply this ratio to the corresponding bin in
the UV continuum slope and determine whether this maximum percentage
can be further constrained by a new fraction $f_{\beta}$.  Finally, we
multiply the old GC percentage by $f_{\beta}$ to obtain our new
constraint.  In Figure~\ref{fig:gc3}\ (right), $\beta_{eff}$ is compared to the
the current values of the UV continuum slope.  Because of the limited
range of the UV continuum slope at the faint end we extrapolate the
least squares fit yet remain conservative in constraining $f_{\beta}$
estimates.  When comparing $\beta_{eff}$ to the known UV continuum
slope with the Power Law and Gaussian models where GCs form in groups
of ten, we can constrain the maximum percentage of GC formation to a
higher degree of accuracy by noting where $\beta_{eff}$ deviates from
the error bars recorded by \cite{Bouwensetal2009}.

\subsection{Unresolved Proto-GC Systems}
\begin{figure*}
\centerline{\epsfig{figure=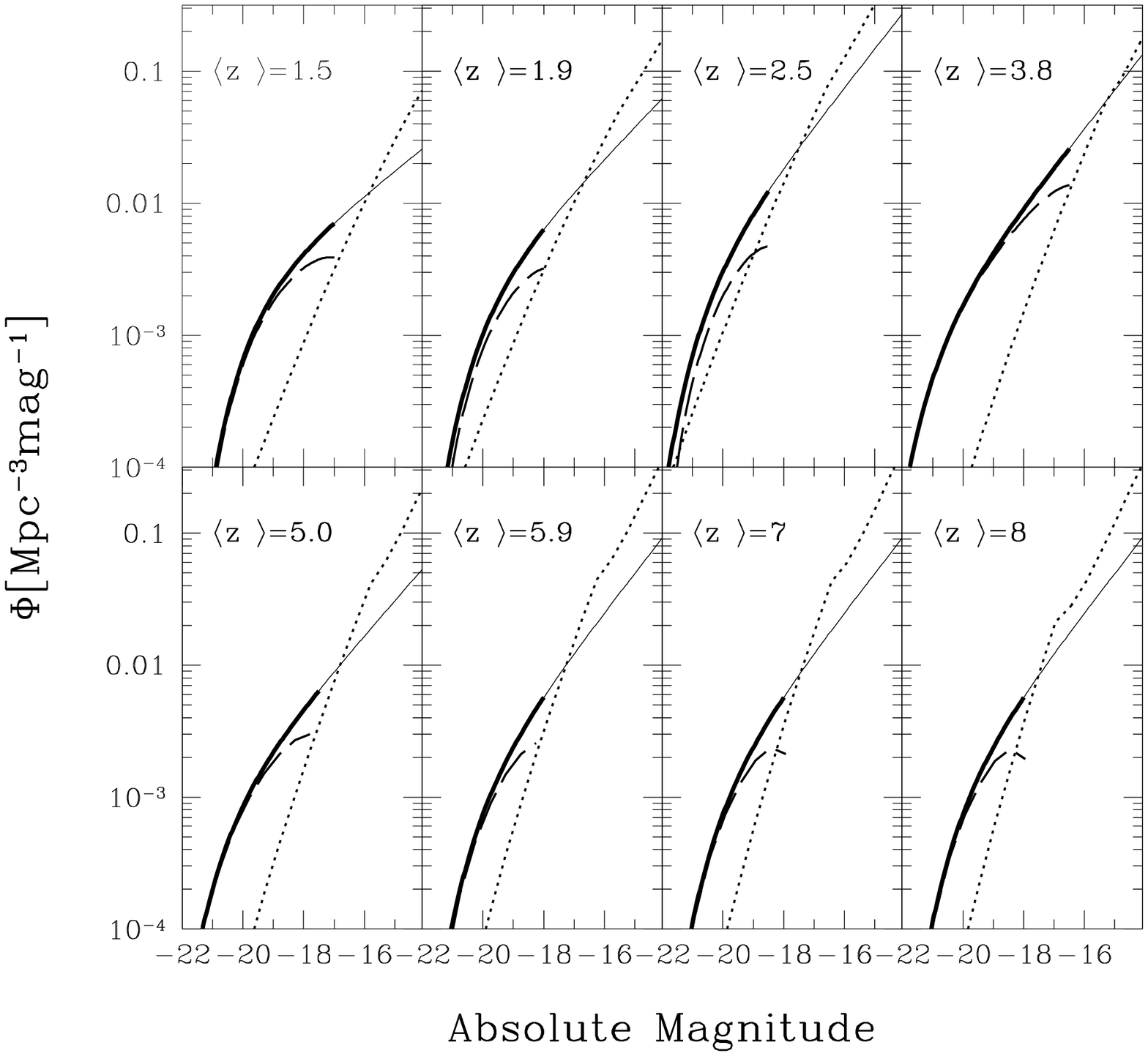, width=8.5cm}\epsfig{figure=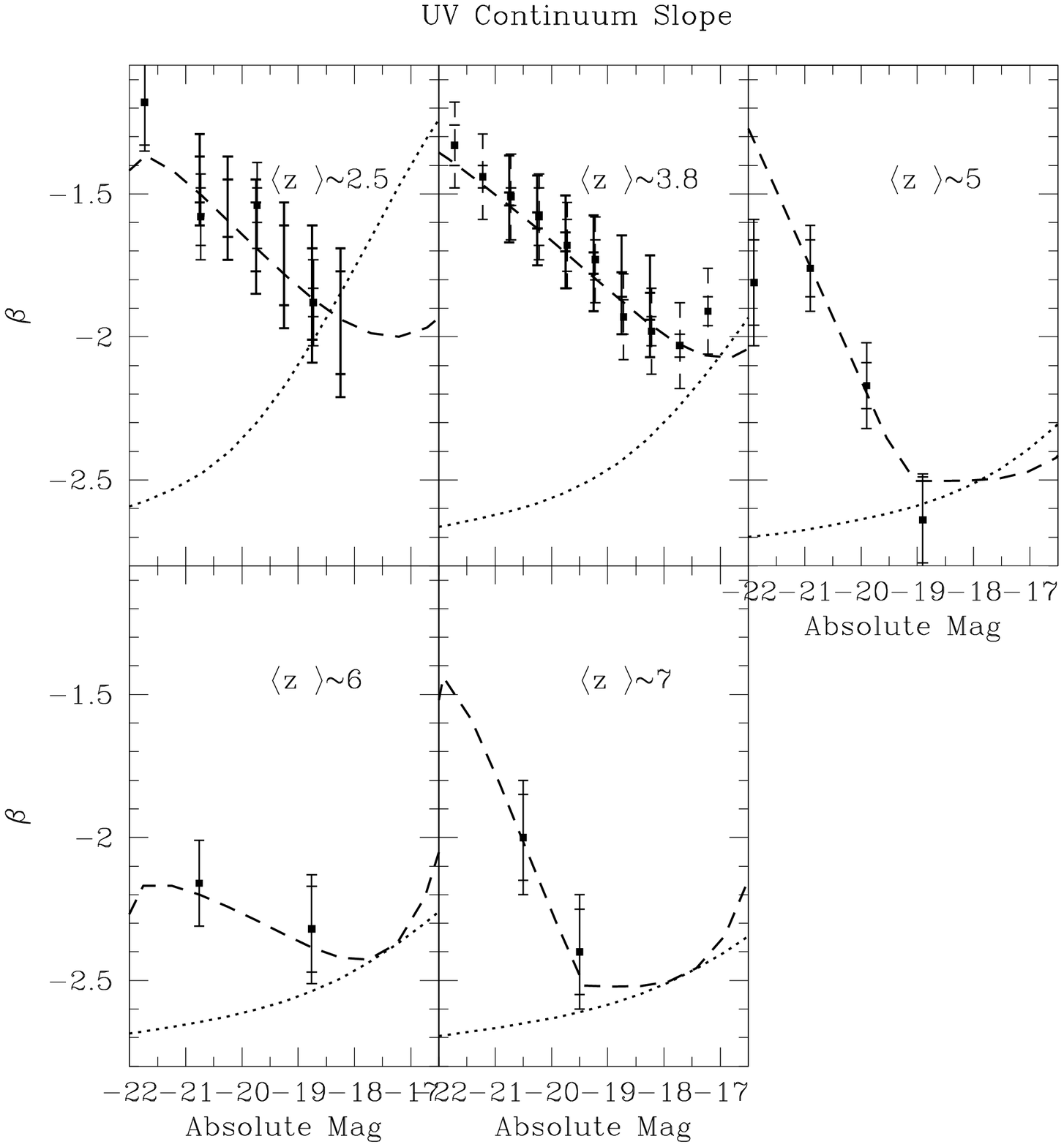, width=8.5cm}}
\caption{The left and the right panels are analogous to the ones in
  Figure~\ref{fig:gc3} but for systems of proto-GCs allowed to form in
  halos with virial temperature $T_{vir}>5 \times 10^{4}$~K. The long
  dashed lines show the difference between the observed LFs and the
  proto-GCs LFs.}
\label{fig:gc4}
\end{figure*}

The second approach considers that more than one proto-GC forms in
each host galaxy, and as discussed in Section~\ref{sec:images},
observations measure the integrated light from the proto-GC system. In
this second case, we do not need to know the GCIMF, but the efficiency
of proto-GC formation as a function of the mass of the host galaxy. We
build on the results from \cite{Puzia} that the mass in GCs $M_{gc,
  tot}$ in a large sample of local dwarf galaxies is proportional to
the total host halo mass $M_{dm}$ within the luminous radius of the
galaxy: $M_{gc, tot}(z=0)= \eta_{dm}(z=0) M_{dm}(z=0)$ with
$\eta_{dm}(z=0)=5.5 \times 10^{-5}$. However, this relationship is
only measured at $z=0$ and does not take into account evolutionary
effects such as galaxy mergers, proto-GC formation and disruption as a
function of time. For instance, we know that only a fraction
$f_{di}(M_{dm},z)\le 1$ of proto-GCs have survived to the
present. Thus, we assume that 
\begin{equation}
M_{gc, tot}(z)=\eta_{dm}(z) M_{dm}(z),
\label{eq:sys}
\end{equation}
where $\eta_{dm}(z)$ is the parameter we wish to constrain.

Modeling GC systems in dark matter halos is somewhat model dependent
as the shape of the luminosity function depends on the range of halo
masses in which proto-GCs form. However, we find that the only
parameter that significantly affects the luminosity function is the
mass of the smallest dark matter halos allowed to form GCs. Our model
assumptions are as follows. We assume a constant efficiency of GC
formation as a function of the halo mass $\eta_{dm}(z)$ for
$M^{min}_{dm}<M_{dm}<M^{max}_{dm}$, and no globular cluster formation outside
this halo mass range. We use the Press-Schechter formalism to obtain
the mass function of dark matter halos as a function of redshift
$dn_{halo}/dM_{dm}$, then convert dark halo masses to the masses of
proto-GCs systems using Equation~(\ref{eq:sys}). Finally, we calculate
the luminosity function by converting the total proto-GC mass to
luminosity using Starburst~99 as described in \S~\ref{sec:single}. As
discussed in the previous here we also include the effect of aged GCs
that form within the redshift bin under consideration and in the
higher redshift bins (note that at high-z there is little time for GCs
to age significantly and become redder).

The proto-GC LF is not sensitive to $M^{max}_{dm}$, that we hence
assume larger than the exponential cut-off mass-scale in the
Press-Schechter mass function. We instead find that the results are
sensitive to $M^{min}_{dm}$, or equivalently to the virial temperature
of the halo $T^{min}_{vir}$. Hence, we explore the results assuming 4
different values of $T^{min}_{vir}=5\times10^4, 8\times 10^4, 10^5$~K
and $1.5\times 10^{5}$~K.  In each model we vary the value of
$\eta_{dm}(z)$ to be consistent with observations of the LFs and
colors in each redshift bin, similarly to what was done for single
proto-GCs. The main different here is that varying $\eta_{dm}$ shifts
the proto-GCs LF horizontally along the magnitude axis, while for
single proto-GCs the free parameter $f_{gc}$ had an effect on the
overall normalization of the LF (shift along the y-axis). 

Figure~\ref{fig:gc4} (left) shows the LF for proto-GCs systems
(dotted line) compared to observed deep field galaxies LFs (solid
lines) in different redshift bins. In each panel, corresponding to a
different redshift, we have used a value of $\eta_{dm}(z)$ that
maximizes the contribution of the proto-GC luminosity function to the
observed LF remaining consistent with observations.  We also repeat
the same procedure described in Section~\ref{sec:slopes} to further
constrain $\eta_{dm}(z)$ from the UV colors of observed deep field
galaxies. Figure~\ref{fig:gc4} (right) shows the UV colors of proto-GCs
systems, $\beta_{gc}$ (dotted lines) and the effective galaxies UV
colors, $\beta_{eff}$ (dashed lines), compared to observations (solid
lines and points with errorbars). The curves in the figure are for the
largest values of $\eta_{dm}(z)$ consistent with observations.

The values of $\eta_{dm}(z)$ constrained by the data allows us to
calculate $\rho_{gc}(z)=\overline \rho_b \eta_{dm}\omega_{halos}$,
where $\omega_{halos}$ is the fraction of collapsed dark matter in
halos with $M^{min}_{dm}<M_{dm}<M^{max}_{dm}$ for each observed
redshift bin of depth $\Delta t(z)$. From $\rho_{gc}$ we derive the
upper limits on the proto-GC formation rate $\dot \rho_{gc}(z) \approx
\rho_{gc}(z)/\Delta t(z)$ and hence upper limits on $f_{gc}(z)$. The
results are shown in Table~\ref{tab:one} for our four model assumptions
and are summarized in Figure~\ref{fig:gc5}\ (left).

\section{Results and Discussion}\label{sec:res}

\begin{table*}
\label{tab:one}
\centering
\begin{minipage}{140mm}
\caption{Constraints on globular clusters formation efficiency from LFs and UV colors. For proto-GCs systems we have assumed $T_{vir}>5\times 10^4$~K.}
\begin{tabular}{@{}lccccccc@{}}
\hline
 & & & \multicolumn{2}{c}{$f_{gc}$ Single GCs} & & \multicolumn{2}{c}{$\eta_{dm}(z)/5.5\times 10^{-5}$ Systems of GCs} \\ 
\cline{4-5} \cline{7-8} \\ 
$\langle z \rangle$ & t[Gyr] & dt[Gyr] & LF   & LF+color & & LF  & LF+color \\
\hline
1.5 & 9.320 & 0.99 & 75\% & -- & \phn & 0.23 & -- \\ 
1.9 & 10.155 & 0.92 & 100\% & -- & \phn & 0.79 & -- \\ 
2.5 & 10.999 & 0.56 & 100\% & 100\% & \phn & 2.56 & 0.77 \\ 
3.8 & 11.995 & 0.36 & 55\% & 30\% & \phn & 1.00 &  1.00 \\ 
5.0 & 12.469 & 0.28 & 35\% & 21\% & \phn & 2.08 &  1.04 \\
5.9 & 12.695 & 0.17 & 40\% & 24\% & \phn & 4.08 &  4.08 \\
6.8 & 12.858 & 0.13 & 28\% & 11\% & \phn & 7.38 &  7.38\\
8.0 & 13.014 & 0.14 & 15\% & -- & \phn &  14.46 & -- \\ 
\hline 
\end{tabular}
\end{minipage}
\end{table*}

\begin{table*}
\label{tab:two}
\centering
\begin{minipage}{140mm}
\caption{Upper limits on globular clusters star formation rate ($M_{\odot}/yr/Mpc^3$)} 
\begin{tabular}{@{}lcccc@{}}
\hline
$z$  & $T_{vir}>1.5\times10^5K$ & $T_{vir}>1.0\times10^5K$ & $T_{vir}>8.0\times10^4K$ & $T_{vir}>5.0\times10^4K$ \\
\hline
4.25 & 0.0063 & 0.0061 & 0.0055 & 0.0051 \\
4.75 & 0.0048 & 0.0050 & 0.0035 & 0.0032 \\
5.25 & 0.0042 & 0.0067 & 0.0052 & 0.0073 \\
5.75 & 0.0050 & 0.0100 & 0.0095 & 0.0151 \\
6.25 & 0.0061 & 0.0105 & 0.0118 & 0.0181 \\
6.75 & 0.0069 & 0.0082 & 0.0112 & 0.0162 \\
7.25 & 0.0067 & 0.0060 & 0.0095 & 0.0144 \\
7.75 & 0.0055 & 0.0051 & 0.0075 & 0.0146 \\
8.25 & 0.0041 & 0.0046 & 0.0054 & 0.0155 \\
8.75 & 0.0026 & 0.0041 & 0.0033 & 0.0164 \\
\hline
\end{tabular}
\end{minipage} 
\end{table*}

\begin{figure*}
\centerline{\epsfig{figure=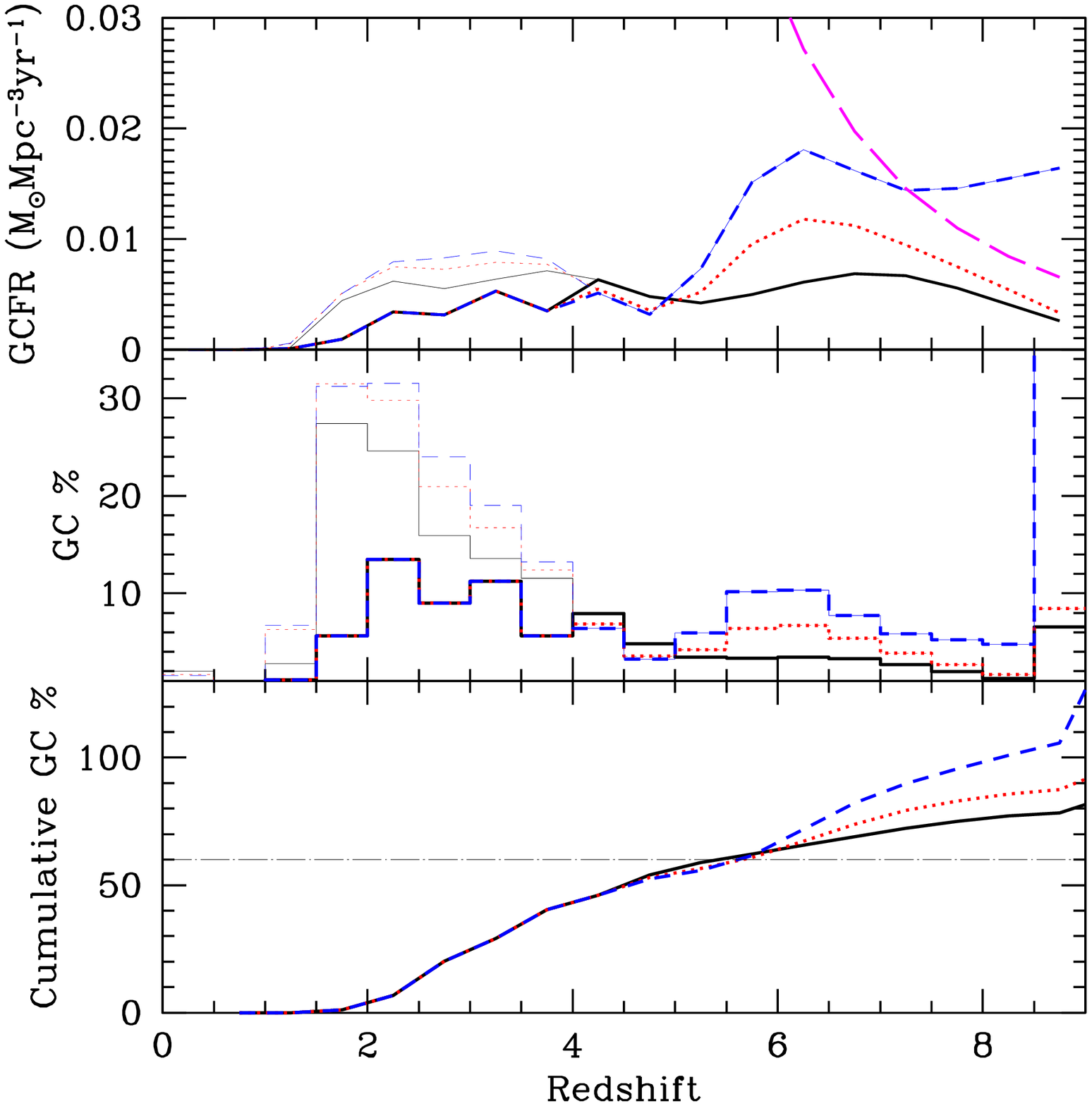, width=8.5cm}\epsfig{figure=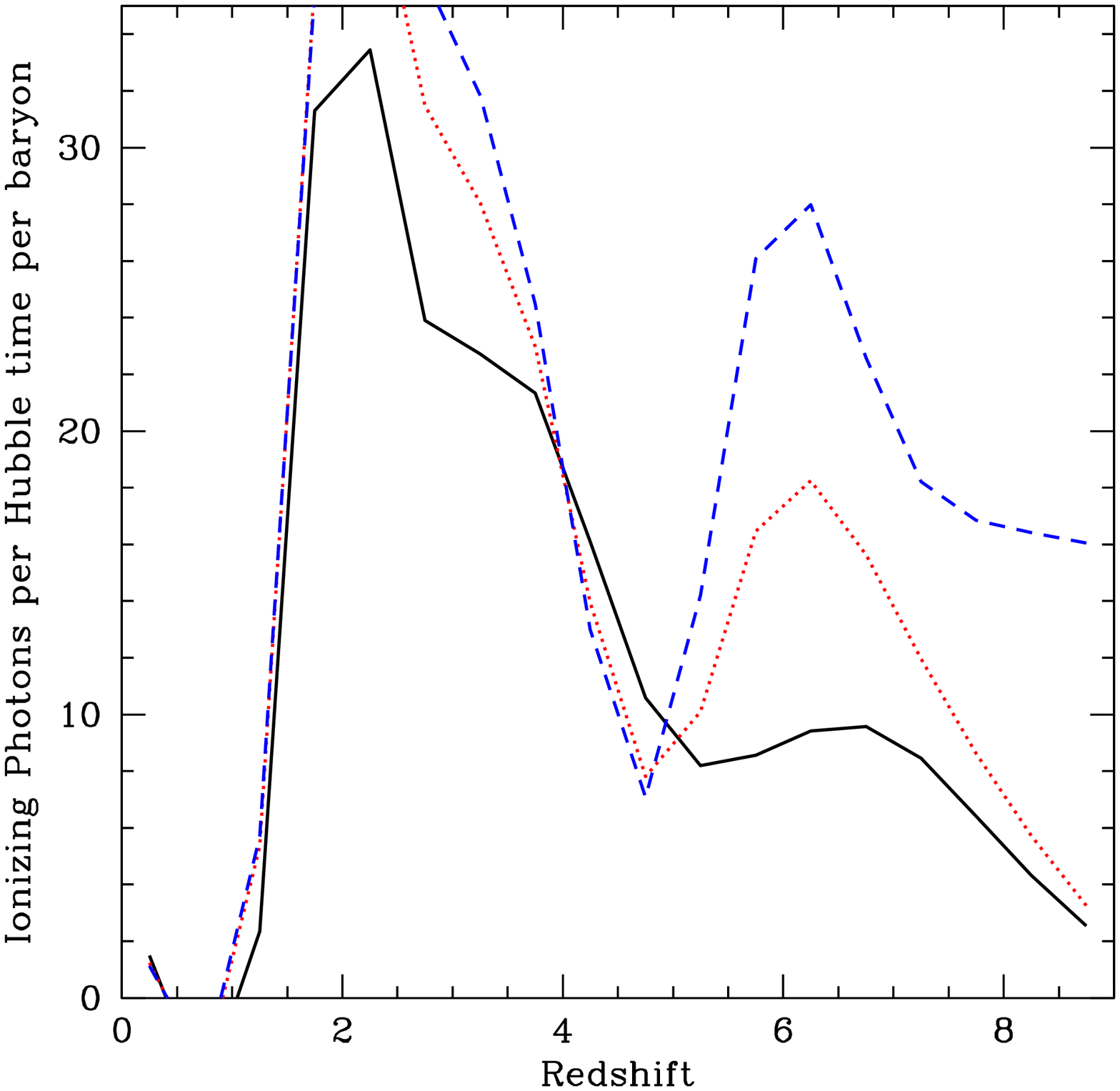, width=8.5cm}}
\caption{{\it Left.} Upper limits on the GC formation rate as a
  function of redshift from LF and color constraints for systems of GCs
  (see text). The top panel shows the GC formation rate compared to the observed SFR (long-dashed line), the middle
  panel shows the distribution of formation redshifts of GCs and the bottom
  panel shows the cumulative distribution. The different lines correspond to
  different assumptions on the minimum virial temperature the host
  halos in which GC systems are allowed to form: $T_{vir}>5\times
  10^4$~K (dashed line); $T_{vir}>8\times 10^4$~K (dotted line)
  $T_{vir}>1.5 \times 10^5$~K (solid line). The thick lines include
  the constraints at low-redshift from estimated GCs ages. {\it
    Right.} Upper limits on the ionization rate per Hubble time per
  baryon from GCs. The lines correspond to the GCs formation rates as
  in the left panel.}
\label{fig:gc5}
\end{figure*}

The upper limits on the proto-GC formation rate from the Hubble deep
fields are summarized in Table~\ref{tab:two} and in Figure~\ref{fig:gc5}. The
table shows the constrains from the observed LFs and
UV colors at $1500$~\AA\ (parameter $\beta$) of high-z galaxies, for both single
proto-GCs and systems of proto-GCs.

Strong constraints are possible because the proto-GC LFs have a
faint-end slope generally steeper than observed, and their spectral
slope at $1500$~\AA\ is typically bluer than observed ($\beta \sim
-2.7$). This is because the instantaneous burst of star formation of
proto-GCs and the subsequent rapid time evolution of their luminosity
from bright to faint, steepens the LF with respect to the intrinsic
slope of the GCIMF (for single proto-GCs).  However, for systems of
proto-GCs the faint-end slope of the LF depends on the slope of the
halo mass function, and thus depends on the masses of the proto-GC
host halos and their redshifts of virialization. This is why, for GC
systems, the limits are sensitive to the assumed minimum mass of halos
that can host proto-GCs.

At high-z the limits on the formation of rate of proto-GC systems are
the most stringent. These, in combination with the measured ages of
GCs in the Milky Way that instead constrain best their lower redshift
formation rate, allow us to conclude that there are two distinct
epochs of proto-GC formation: one at $z\sim 2.5$ near the peak of the
global SFR density in the universe, and one before or around the
reionization epoch.  Figure~\ref{fig:gc5} (left) shows the upper
limits on the GCs formation rate (top panel), the proto-GC redshift
distribution (middle panel) and the cumulative proto-GC redshift
distribution (bottom panel) from the combined constraints from the LF
and colors discussed in the previous section. In addition, the
proto-GC redshift distribution and cumulative distributions are
constrained\footnote{This constrain implicitly assumes that Milky Way
  GCs accurately describe the age distribution of the GC population in
  the rest of the Universe.} at redshifts $z<3$ by the measured ages
of Milky Way GCs from \cite{ForbesBridges2010}.  The lines in
Figure~\ref{fig:gc5} (left) correspond to different assumptions on the
minimum mass of the halos in which proto-GC systems form: $T_{vir}>5\times 10^4$~K (dashed line),
$T_{vir}>8\times 10^4$~K (dotted line); and $T_{vir}>1.5\times 10^5$~K
(solid line).  The
limits on the GCs formation rate are independent of the assumed
fiducial value for $\rho_{gc}$, while the redshift distribution and
cumulative redshift distribution of GC are normalized to $\rho_{gc}$
(that is somewhat uncertain as it depends on $f_{di}$). Because of the
uncertainty on $f_{\rm di}$ and because the lines represent upper
limits, the cumulative GC distribution may not be normalized to
unity.

Although our results only set upper limits on the GC formation rate,
at redshifts $z\simgt 6$ the LFs and colors of simulated proto-GC
systems are in good agreement with observations, suggesting that a large
fraction of high redshift galaxies have a star formation mode
compatible with proto-GC formation. Moreover, the upper limit on the
cumulative distribution of proto-GCs normalized to $\rho_{gc}$
approaches unity, also suggesting that the upper limits are close to
the actual values of the distribution.  We also note that even a small
fraction (10-20\%) of present-day GCs forming at $z \simgt 6$ accounts
for a significant fraction of the total star formation activity at
high redshift.  Our model shows that if proto-GCs are allowed to form
in halos with $T_{vir} \simgt 5\times 10^4$~K (dashed lines), current
deep fields are missing a significant fraction of the global SFR
because of the steep faint-end slope of the LF.

Finally, we re-asses the possibility that proto-GCs may be dominant
sources of IGM reionization \citep{Ricotti2002}.  The high star
formation efficiency of young proto-GCs and their location in the
outer parts of galaxies both point to a high escape fraction of the
ionizing radiation they emit.  Figure~\ref{fig:gc5} (right) shows the
upper limits on the number ${\cal N}$ of ionizing photons emitted per
Hubble time per baryon by proto-GC formation, assuming
$f_{esc}=1$. The meaning of the lines is the same as in the left
panel. A value of ${\cal N} \sim 5$ is sufficient for reionizing the
IGM and keeping it ionized \citep{Miraldaetal2000}.  This scenario is
especially interesting in light of current studies that are trying to
identify the sources of reionization in deep surveys at $z \sim 7-8$,
finding that the sources of reionization must be faint and thus
relatively numerous \citep{Bouwensetal2010}.

\section{Summary and Conclusions}\label{sec:conc}

Present day GCs are difficult to observe beyond the Milky Way because
their stellar populations are very old and faint. All their bright
massive stars have evolved beyond the main sequence and the surviving
clusters have lost a significant fraction of their lower mass
stars due to dynamical effects (tides, two-body relaxation, dynamical
friction). However, during the first 5-10~Myr after their formation,
proto-GCs were extremely bright, with the most massive GCs
reaching absolute magnitudes of $M_V \approx -19$. This is due to their
high-efficiency of star formation and their relatively large
total mass at formation that can been estimated about 5-10 times
larger than their current stellar mass [although this number is
uncertain, depending on dynamical effects
\citep{OstrickerGnedin1997} as well as the IMF of their stars
\citep{SchaererCharbonnel2011}].

In this paper we have shown that the most massive proto-GCs are
detectable in the HUDF and will be detectable by the JWST up to
redshift $z\sim 8$, when imaged at their peak luminosity. In addition,
since GCs are found with a high specific frequency in present day dwarf
galaxies (5-10 within of a few kpc from the dwarf center), and we
expect their formation to be synchronized to within $\sim 10$~Myrs,
such proto-GC systems appear a poorly resolved high redshift galaxy
with multiple knots of star formation. The short dynamical time scales
in primordial dwarf galaxies support this scenario. However, because
proto-GCs maintain their peak luminosity for a very small fraction of
the Hubble time, their numbers are small within the limited field of
view of deep HST and JWST images, especially if a large fraction of
them form at low redshifts when the Hubble time is large.  But if a
non-negligible fraction of today's GCs have formed at redshifts $z>3$,
they are detectable in the HUDF and will contribute a substantial
fraction of the high redshift stellar populations and to IGM
reionization, as first noted by \cite{Ricotti2002}.

Our present work has shown that only a small fraction of the
present-day GCs populations could have formed in the redshift range
$3<z<5$, because otherwise they would affect significantly and be
inconsistent with the observed LFs and the colors at $1500$~\AA\ of
high-z galaxies in the HUDF. In addition, given the age estimates of
Milky Way GCs, a significant old population that formed at $z>3$ must
exist (even if we take into account the large errors on their age
estimate from color-magnitude methods).  We conclude that the bulk of
the old GC population with estimated ages $\simgt 12$~Gyr, roughly
$50\%$ of the total, formed in the most massive dwarf galaxies at
redshifts $z\simgt 6$ and that proto-GC formation, an important mode of
star formation in some primordial dwarf galaxies, likely dominated
the reionization process.

Although our study only sets upper limits on the formation rate of
proto-GCs, the steep faint-end slope of the LFs and the blue colors of
$z\sim 7-8$ galaxy candidates in the HUDF are consistent with a
significant fraction of their stellar populations being proto-GCs. The
simulated images of systems of proto-GCs forming in $z\sim 7-8$ dwarf
galaxies present remarkable morphological similarities with some of
the observed candidates. JWST will be able to detect and image this
population that is likely responsible for the reionization of the
universe, to higher redshifts and fainter magnitudes.

Finally, we speculate that if GC formation was a dominant mode of star
formation in some dwarf galaxies, this may have implications to
understanding some unsolved issues regarding the Milky Way
satellites. \cite{Bovill2009} have pointed out that many of the
ultra-faint Milky Way satellites have properties that are consistent
with being the fossil remnants of the first dwarf galaxies formed
before reionization. But they also noted that a problem remains
regarding an apparent lack of massive and bright satellites that
cannot be reconciled by simply populating the most massive sub-halos
with luminous satellites using monotonic luminosity-to-mass matching
methods \citep{BovillR2011a, BovillR2011b}. \cite{Boylan-Kolchin2011} have also
suggested a possible problem with the density of the most massive
Milky Way satellites that seems inconsistent with dark matter
$\Lambda$CDM simulations (neglecting baryonic physics). Solutions to
this second problem include SN feedback processes that may reduce the
central density of satellites, and statistical arguments involving
a better knowledge of the mass and circular velocity of the Milky
Way. However, these two solutions would not solve the problem emerging
from \cite{BovillR2011b} work.

A solution that would reconcile both results is to assume that
classical dwarfs do not reside in the most massive dark matter
satellites. This is possible only if the most massive satellites of
the Milky Way are darker than some of the least massive (\ie, a
non-monotonic mass-to-light ratio). There are two ways to achieve
this result: i) suppress star formation preferentially in the most
massive satellite progenitors, with a feedback process such as
photoheating by AGN etc.; or ii) remove most of its stars without
removing most of the dark matter. This second proposal seems
implausible if stars are concentrated at the center of the dark matter
halo. However, if star formation in some of the most massive surviving
Milky Way satellites was dominated by proto-GC formation in the
outskirts of their dark matter halo (as observed in nearby dwarf
galaxies), tidal stripping of their GC systems may significantly
reduce their luminosity without destroying the dark halo, thus
explaining their apparent darkness. Examples of galaxies with stellar
populations clearly dominated by GCs have not been identified in
isolation in the local universe, but their formation might be biased,
such that today they have mostly merged into massive galaxies. Another
possibility is that isolated massive dwarfs were once dominated by GC
systems but have accreted gas and formed new stars at later times
\citep{Ricotti2009}, while the ones that merged into the Milky Way did
not experience further star formation.

\bibliographystyle{./apj}
\bibliography{./hbib}

\label{lastpage}
\end{document}